\documentclass[a4paper]{panl}
\usepackage{cite}
\usepackage{wrapfig}
\usepackage{graphicx}
\usepackage{amssymb}
\usepackage{amsfonts}
\usepackage{amsmath}
\usepackage{longtable}
\usepackage{rotating}
\usepackage{lscape}
\usepackage{epsfig}
\usepackage{multirow}

\originalTeX
\begin{document}

\title{Consistent  Lagrangians for irreducible  interacting higher-spin fields with holonomic constraints
} \maketitle \setcounter{footnote}{0} \authors{I.L.
Buchbinder$^{a,b,c}$\footnote{E-mail: joseph@tspu.edu.ru},
A.A.\,Reshetnyak$^{a,d}$\footnote{E-mail: reshet@tspu.edu.ru}}
\from{$^{a}$\, Center for Theoretical Physics, Tomsk State
Pedagogical University, Tomsk, Russia}
\from{$^{b}$\,Bogoliubov Laboratory of Theoretical
Physics, JINR, Dubna,  Russia}
\from{$^{c}$\,Tomsk State University of Control
Systems and Radioelectronics, Tomsk, Russia}
\from{$^{d}$\,National
Research Tomsk Polytechnic University, Tomsk, Russia}
\begin{abstract}

\vspace{0.2cm}

We study the aspects of constructing the interactions for the higher
spin fields in the framework of BRST approach. The main object of
such an approach is BRST operator acting in the appropriate Fock
space and building on the base of constraints that define the
irreducible higher spin representations. In its turn, the constrains
are divided into differential and purely algebraic or holonomic. The
necessary and sufficient conditions to derive the consistent
Lagrangian formulations for irreducible interacting higher-spin
fields within approach with incomplete BRST operator, where the
algebraic constraints are not included into definition of the BRST
operator but imposed "by hands" on the field and gauge parameter vectors,
are considered. It is shown that in addition to that such
constraints should (anti)commute with the BRST operator and
annihilate the fields and gauge parameters Fock space vectors, they
must form the Abelian (super)algebra both with the BRST operator
above and with operators of cubic, quartic and etc. vertices. Only
under the above conditions, the formulations with complete and
incomplete BRST charges turn out to be equivalent and yield to the
same interaction vertices in terms of irreducible fields.

\end{abstract}
\vspace*{6pt}


\label{sec:intro}
\section*{Introduction}

Higher spin field theory (for a review see, e.g., the recent papers
\cite{Snowmass}, \cite{Ponomarev} and the references therein) can be
treated as a polygon for searching the new approaches to unification
of fundamental interactions beyond the Standard Model. Given the
need to use quantum field methods to describe physical processes
with elementary particles of higher spins, one of the most important
tasks in the theory of higher spin fields is the development of a
Lagrangian formulation of such fields, taking into account their
interactions.

There are two well-adopted  ways to construct gauge-invariant
Lagrangians for higher spin fields both in metric- and in frame-like
formalisms. In the  first one, the fields remains by unconstrained,
i.e. all the conditions (d'Alamberti\-an, divergentless,  traceless),
which select the irreducible representation of the Poincar\'e group
with given mass and spin are included on equal footing to get
Lagrangian. In the second one, part of these constraints, usually
related with traces, are not used at Lagrangian construction but are
consistently imposed "by hands"  on the equations of motion which
follow from such  Lagrangian. On the level of free fields there
exist two efficient approaches to realize these aims, known as the
BRST approaches respectively with complete (see e.g.
\cite{BPT})\footnote{At present, there is an extensive literature
devoted to the BRST approach for free fields of higher spins; here
we only cite one of the first works in this direction} and
incomplete \cite{Alkalaev} BRST operators which Lagrangian
formulations for the same higher-spin field in $d$-dimensional
Minkowski space-time was shown to be equivalent in
\cite{Reshetnyak_con}.

Let us now turn to the higher spin theories with cubic
interaction (see e.g. \cite{Manvelyan}, \cite{Manvelyan1},
\cite{Joung}, \cite{frame-like1}, \cite{BRST-BV3},
\cite{frame-like2}, \cite{BKTW} and the references therein).
Classification of the higher spin cubic vertices was given in
light-cone formalism by Metsaev \cite{Metsaev0512}. The
covariantization of this result within incomplete BRST approach
\cite{BRST-BV3} works perfectly for reducible interacting higher
spin fields, where the traceless conditions are not included into
BRST operator and therefore should be imposed "by hands" on the
fields. It is clear that such an approach is not completely
Lagrangian and it is not evident that this procedure is equivalent
to approach in terms of complete BRST operator based on all
appropriate constraints of the theory. In fact, the incomplete BRST
approach does not take into account the effect of holonomic
constraints with traces on the vertex's structure. This circumstance
destroys tracelessness both for the deformed equations of motion and
for gauge transformations, which leads to an increase in the number
physical degrees of freedom in comparison with  corresponding free
theory.

In the recent papers \cite{BRcub}, \cite{Rcubmasless},
\cite{BRcubmass} (see as well \cite{BKStwis} for $d=4$)  the
solutions for Lagrangian cubic vertices  has been derived for
interacting unconstrained massless and massive totally-symmetric
fields in Minkowski spaces within approach with complete BRST
operator (see, e.g. \cite{mixResh} and the references therein). The
solutions are modified,  have additional terms as compared to
\cite{BRST-BV3}, automatically preserve the true number of physical
degrees of freedom when passing to interacting case, hence providing
construction of the covariant cubic vertices for irreducible
higher-spin fields.

Thus, we face the problem which additional conditions should be
imposed on the theory in the incomplete BRST approach in order to
this approach yields the same higher spin interaction vertices as
the complete BRST approach. The first main purpose of the paper is
to develop the consistent deformation procedure for the approach
with incomplete BRST operator to get cubic, quartic and so on
vertices for irreducible totally-symmetric higher integer spin
fields on $d$-dimensional Minkowski space. Second, we intend to
establish interrelation among Lagrangian formulation  for the fields
above obtained from  methods with complete and incomplete BRST
operators.

The paper is organized as follows.  In
Section~\ref{sec:preparation}, the results of the BRST construction
using a incomplete BRST operator are presented. In
Section~\ref{sec:consistent}, we derive necessary and sufficient
conditions for the incomplete BRST operator, BRST-extended holonomic
constraint, vertex operators to have non-contradictory Lagrangian
dynamics. Section~\ref{sec:completeBRST} deals with relation of the
interacting Lagrangian formulations deriving on a base of complete
and incomplete BRST operators. Conclusion resumes the results.

We use the conventions of \cite{BRcub}, \cite{BRcubmass} :
$\eta_{\mu\nu} = diag (+, -,...,-)$ for a metric tensor with Lorentz
indices $\mu, \nu = 0 ,1,...,d-1$, the notation
$\epsilon(F)$,$gh(F)$, $[x]$, $[H,\,G\}$, $\theta_{m,0}$, $(s)_k$
for the respective Grassmann parity and ghost number of a
homogeneous quantity $F$, as well as the integer part of a number
$x$,  the supercommutator of quantities $H,G$,  Heaviside
$\theta$-symbol to be equal to $1(0)$ when $m>0(m\leq 0)$ and for
integer-valued vector $(s_1,...,s_k)$.

\section{Constrained BRST approach for free higher-spin fields}
\label{sec:preparation}

The unitary massless (massive) irreducible representations of
Poincar\'e  group with integer helicity  (spin) $s$ may be realized
using the $\mathbb{R}$-valued totally symmetric tensor fields $
\phi_0^{\mu(s)}\equiv \phi_0^{\mu_1...\mu_s}(x)$  included into
basic vectors $|\phi\rangle$ subject to the conditions
\begin{eqnarray}\label{irrepint}
       &\hspace{-0.7em}& \hspace{-0.7em}      \big(g_0 -d/2, l_0,\, l_1,\, l_{11}\big)|\phi\rangle  = (s,0,0,0)|\phi\rangle \ \mathrm{for} \  |\phi\rangle  =  \sum_{s\geq 0}\frac{\imath^s}{s!}\phi_0^{\mu(s)}\prod_{i=1}^s a^+_{\mu_i}|0\rangle.
\end{eqnarray}
The operators  of number particles $g_0$, wave  $l_0,\,$
divergent $l_1,$ traceless $l_{11}$ constraints above  are defined in the Fock space $\mathcal{H}$
with the  oscillators $a_\mu, a^+_\nu$, ($[a_\mu, a^+_\nu]= - \eta_{\mu\nu}$)  as follows
\begin{eqnarray}\label{FVoper}
&\hspace{-0.7em} &\hspace{-0.7em}  \big( g_0,\, l_0,\, l_1,\, l_{11}\big) = \big( -\frac{1}{2}\big\{a^+_{\mu},\, a^{\mu}\big\},\,\partial^\nu\partial_\nu+ m^2 ,\, - \imath a^\nu  \partial_\nu ,\, \frac{1}{2}a^\mu a_\mu\big).\nonumber
\end{eqnarray}
The free dynamics of the field with
definite spin $s$ in the framework of constrained  BRST approach  is described by the irreducible gauge theory with the gauge invariant action given on the configuration space $M^{(s)}_{c|cl}$ which includes  triplet of  fields $\phi_i^{\mu(s-i)}$ for $i=0,1,2$ (for massive case 3 sets of fields $\phi_{i|k}^{\mu(s-i-k)}$ at $k=0,1,...,s-i$) incorporated into the vector $|\chi_c\rangle_s$
\begin{eqnarray}
\label{PhysStatetot} \mathcal{S}^m_{0|s}[\Phi_i]=
\mathcal{S}^m_{0|s}[|\chi_c\rangle_s] = \hspace{-0.3em} \int\hspace{-0.2em} d\eta_0 {}_s\langle\chi_c|
Q_c|\chi_c\rangle_s, \ \ \delta\big(|\chi_c\rangle_s,\, |\Lambda_c\rangle_s\big)  =  \big(Q_c|\Lambda_c\rangle_s , \,0\big)
\end{eqnarray}
and subject to the additional off-shell constraints
\begin{eqnarray}\label{L11}
    && L_{11} \big(|\chi_c\rangle_{s}, |\Lambda_c\rangle_{s}\big) =0.
\end{eqnarray}
Here $\eta_0$, $Q_c$, $L_{11}$  and $ |\Lambda_c\rangle_s$  be respectively a zero-mode ghost, nilpotent incomplete BRST operator , BRST-extended  traceless constraint aad  parameter of  gauge transformations  having the form
\begin{eqnarray}
&& \big({Q}_c,\, L_{11} \big)\  =\ \big(
\eta_0l_0+\eta_1^+\check{l}_1+\check{l}_1^{+}\eta_1
 + {\imath}\eta_1^+\eta_1{\cal{}P}_0,\ l_{11}-\frac{1}{2}d^2+\eta_{1} \mathcal{P}_{1} \big)
\label{Qctotsym}
\end{eqnarray}
(for $\big(\check{l}_1, \check{l}_1^{+}\big) = \big({l}_1+md, {l}_1^{+}+md^+\big) $ with additional bosonic oscillators $d, d^+$:  $[d, d^+]=1$).
The Grassmann-odd  ghost oscillators $\eta_0,  \mathcal{P}_{0},  \eta_1,  \mathcal{P}_{1}^+, \eta_1^+,  \mathcal{P}_{1}$ correspond to the system of first-class differential constraints $l_0, \check{l}_1,\check{l}_1^+$ with algebra $[\check{l}_1,\, \check{l}_1^+]= l_0$  and obeys to the non-vanishing anticommutator relations
    \begin{equation}\label{ghanticomm}
  \{\eta_0, \mathcal{P}_0\}= \imath,\ \  \{\eta_1, \mathcal{P}_1^+\}=1,  \ \ (\epsilon, gh)\eta_{...} = (\epsilon,  - gh)\mathcal{P}_{...}=(1,1).
\end{equation}
The label "$s$" at field and gauge parameter vectors
\begin{eqnarray}
\label{extconstsp5}
 &\hspace{-0.5em} &\hspace{-0.5em}  |\chi_c\rangle_{s} = |\Phi_0\rangle_{s}-\mathcal{P}_1^{+}\big(\eta_0|\Phi_1\rangle_{s-1}+\eta_1^{+}|\Phi_2\rangle_{s-2}\big),  \ \  |\Lambda_c\rangle_s = \mathcal{P}_1^{+}|\Xi\rangle_{s-1},\\
 &\hspace{-0.5em} &\hspace{-0.5em}  |\Phi_i\rangle_{s-i}= \sum_{k=0}^{s-i}\frac{(d^+)^k}{k!}|\phi_{i|k}(a^+)\rangle_{s-k-i},\ \mathrm{for} \    |\phi_{0|0}(a^+)\rangle_{s}\equiv |\phi\rangle_s
\end{eqnarray}
means that these vectors are proper eigen-vectors for the spin operator with definite spin value $s$
\begin{eqnarray}
&&   \sigma_{c} \Big(|\chi_c\rangle_{s}, |\Lambda_c\rangle_{s}\Big) = \Big(s-1+\frac{d+\theta_{m,0}}{2}\Big)\Big(|\chi_c\rangle_{s}, |\Lambda_c\rangle_{s}\Big), \label{s11} \\
 &&  {\sigma}_c   =   g_0 + \theta_{m,0} d^{+}d^{} + \frac{1}{2}+ \eta_1^{+}\mathcal{P}_{1}
-\eta_1\mathcal{P}_{1}^{+}.
\end{eqnarray}
The incomplete
 BRST operator $Q_c$ forms with  traceless  constraint and constrained spin operator closed superalgebra \cite{Reshetnyak_con}:
   \begin{eqnarray}
 &&   (Q_c)^2 \ = \  [ Q_c, \,  L_{11}\} \ =\ [ Q_c, \,  \sigma_{c}\}= 0, \ \ [\sigma_{c},\,  L_{11}\}= -2 L_{11}. \label{QsL11}
\end{eqnarray}
Note, that both the equations of motion, $Q_c|\chi_c\rangle_{s} =0$, and,  that any field representative ($\big|\widetilde{\chi}_c\rangle_{s}$) from the gauge orbit
\begin{equation}\label{goc}
\mathcal{O}_{0|\chi_c} = \big\{\big|\widetilde{\chi}_c\rangle_{s} \big|\, \  \big|\widetilde{\chi}_c\rangle_{s} =\big|{\chi}_c\rangle_{s} +  Q_c \big|\Lambda_c\rangle_{s}, \, \forall \big|\Lambda_c\rangle_{s}\big\}
\end{equation}
 remains by traceless if the field $\big|{\chi}_c\rangle$ and  gauge parameter $ \big|\Lambda_c\rangle$  are traceless because of the commutation of
 $L_{11}$ with $Q_c$.

One can easily show that after resolving the traceless constraints
and eliminating the auxiliary fields from equations of motion, the
theory under consideration is reduced to Fronsdal \cite{Fronsdal} (Singh-Hagen for massive case, see \cite{BRcubmass}) form in terms of
totally symmetric double traceless tensor field $\phi_{\mu(s)}$  and
traceless gauge parameter $\xi_{\mu(s-1)}$ (without gauge invariance for massive field).

\section{Consistent deformation procedure  for interacting  higher-spin fields}
\label{sec:consistent}

To  include the  interaction   we introduce $k$, $k\geq 3$ copies of Lagrangian formulations with vectors
$|\chi^{(j)}_c\rangle_{s_j}$, gauge parameters
$|\Lambda^{(j)}_c\rangle_{s_j}$   with
corresponding vacuum vectors $|0\rangle^j$ and oscillators for
$j=1,...,k$. It permits to  define the deformed  action and gauge transformations up to $p$-tic vertices, $p=3,4,...,e$  in  powers of deformation  constant $g$, starting from sum of $k$ copies of Lagrangian formulations for free higher-spin fields and then from cubic, quartic and so on vertices:
\begin{eqnarray}\label{S[e]}
  && S^{(m)_k}_{[e]|(s)_k}[(\chi_c)_k] \  = \  \sum_{j=1}^{k} {S}^{m_j}_{0|s_j}[\chi^{(j)}_c]   + \sum_{h=1}^e g^h S^{(m)_k}_{h|(s)_k}[(\chi_c)_k],
  \end{eqnarray}
  where
\begin{eqnarray}\label{S[3]}
&\hspace{-0.5em}&\hspace{-0.5em}  S^{(m)_k}_{1|(s)_k}[(\chi_c)_k] =   \sum_{1\leq i_1<i_2<i_3\leq k} \hspace{-1.0em} \int \prod_{j=1}^3 d\eta^{(i_j)}_0  \Big( {}_{s_{i_j}}\langle \chi^{(i_j)}_c
  \big|  V^{(3)}_c\rangle^{(m)_{(i)_3}}_{(s)_{(i)_3}}+h.c. \Big)  , \\
  \label{S[4]}
&\hspace{-0.5em}&\hspace{-0.5em}  S^{(m)_k}_{2|(s)_k}[(\chi_c)_k] =   \sum_{1\leq i_1<i_2<i_3<i_4\leq k}\hspace{-1.0em}  \int \prod_{j=1}^4 d\eta^{(i_j)}_0  \Big( {}_{s_{i_j}}\langle \chi^{(i_j)}_c
  \big|  V^{(4)}_c\rangle^{(m)_{(i)_4}}_{(s)_{(i)_4}}+h.c. \Big)  , \\
   && \ldots \ \ldots\ \ldots \ \ldots\ \ldots \ \ldots\ \ldots \ \ldots\ \ldots \ \ldots \ \ldots \ \ldots\ \ldots \ \ldots \nonumber\\
    \label{S[+]}
&\hspace{-0.5em}&\hspace{-0.5em}  S^{(m)_k}_{e|(s)_k}[(\chi_c)_k] =   \sum_{1\leq i_1<i_2<...<i_e\leq k}  \hspace{-1.0em}\int \prod_{j=1}^e d\eta^{(i_j)}_0  \Big( {}_{s_{i_j}}\langle \chi^{(i_j)}_c
  \big|  V^{(e)}_c\rangle^{(m)_{(i)_e}}_{(s)_{(i)_e}}+h.c. \Big) ,
\end{eqnarray}
also for deformed gauge transformations: $\delta_{[e]}  |\chi^{(j)}_c\rangle_{s_j}=(\delta_0+\delta_1+...+\delta_e )|\chi^{(j)}_c\rangle_{s_j} $
\begin{eqnarray}\label{gt1}
&\hspace{-0.5em}&\hspace{-0.5em} \delta_1|\chi^{(j)}_c\rangle_{s_j}\hspace{-0.1em} =  - \hspace{-1.0em}\sum_{1\leq i_1<i_2\leq k}\hspace{-0.5em} \int \hspace{-0.3em}\prod_{j=1}^2 d\eta^{(i_j)}_0  \Big[ {}_{s_{i_1}}\langle \chi^{(\{i_1)}_c
  \big|  {}_{s_{i_2}}\langle \Lambda^{(i_2\})}_c
  \big| \widetilde{V}{}^{(3)}_c\rangle^{(m)_{(i)_2j}}_{(s)_{(i)_2j}}\hspace{-0.1em}+h.c. \Big]\hspace{-0.1em}  , \\
  \label{gt4}
&\hspace{-0.5em}&\hspace{-0.5em}  \delta_2|\chi^{(j)}_c\rangle_{s_j} =  - \sum_{1\leq i_1<i_2<i_3\leq k}\hspace{-1.0em}  \int \prod_{j=1}^3 d\eta^{(i_j)}_0  \Big[  {}_{s_{i_1}}\langle \chi^{(\{i_1)}_c
  \big| {}_{s_{i_2}}\langle \chi^{(i_2)}_c
  \big| {}_{s_{i_3\}}}\langle\Lambda^{(i_3)}_c
  \big| \\
  &\hspace{-0.5em}&\hspace{-0.5em} \phantom{\delta_2|\chi^{(j)}_c\rangle_{s_j}\ \ }\times
   \big| \widetilde{V}{}^{(4)}_c\rangle^{(m)_{(i)_3j}}_{(s)_{(i)_3j}}\big)+h.c. \Big]  , \nonumber \\
   &\hspace{-0.5em}&\hspace{-0.5em} \ldots \ \ldots\ \ldots \ \ldots\ \ldots \ \ldots\ \ldots \ \ldots\ \ldots \ \ldots \ \ldots \ \ldots\ \ldots \ \ldots \nonumber\\
    \label{gte}
&\hspace{-0.5em}&\hspace{-0.5em}  \delta_e|\chi^{(j)}_c\rangle_{s_j} =   -\sum_{1\leq i_1<...<i_{e-1}\leq k} \hspace{-1.0em} \int \prod_{j=1}^{e-1} d\eta^{(i_j)}_0  \Big[{}_{s_{i_1}}\langle \chi^{(\{i_1)}_c
  \big| \ldots {}_{s_{i_{e-2}}}\langle \chi^{(i_{e-2})}_c
  \big| \\
  &\hspace{-0.5em}&\hspace{-0.5em} \phantom{\delta_2|\chi^{(j)}_c\rangle_{s_j}\ \ }\otimes
   {}_{s_{i_{e-1}}}\langle\Lambda^{(i_{e-1}\})}_c
  \big|  \widetilde{V}{}^{(e)}_c\rangle^{(m)_{(i)_{e-1}j}}_{(s)_{(i)_{e-1}j}}+h.c. \Big] , \nonumber
\end{eqnarray}
where we have used the notations $(\chi_c)_k= (\chi^{(1)}_c, \chi^{(2)}_c, ..., \chi^{(k)}_c)$, the symmetrization  of indices $\{i_1,...,i_{e-1}\}$  and the number of different terms in $p$-tic vertex is equal to $k!/((k-p)!p!)$.

In order to the interacting theory constructed from  initial  actions ${S}^{m_j}_{0|s_j}$, $j=1,...,k$ would  preserve the number of physical  degrees of freedom $N_j$ determined by Lagrangian formulations for free higher-spin fields  with spin $s_j$, we demand,  that the sum of all physical  degrees of freedom would be unchangeable, i.e. $\sum_{i}N_I = \mathrm{const}$. This property will be guaranteed, first, if the deformed  action $S^{(m)_k}_{[e]|(s)_k}$ will satisfy to sequence of deformed Noether identities in powers of $g$:
\begin{eqnarray}
    &g^0:&   \delta_0 S^{(m)_k}_{[0]|(s)_k}[(\chi_c)_k] \equiv 0  , \label{g0}  \\
   &g^1:& \delta_0 S^{(m)_k}_{1|(s)_k}[(\chi_c)_k]   + \delta_1 S^{(m)_k}_{[0]|(s)_k}[(\chi_c)_k]  =0   , \label{g1} \\
    &g^2:& \delta_0 S^{(m)_k}_{2|(s)_k}[(\chi_c)_k]+ \delta_1 S^{(m)_k}_{1|(s)_k}[(\chi_c)_k] + \delta_2 S^{(m)_k}_{0|(s)_k}[(\chi_c)_k] = 0 , \label{g2} \\
&& \ldots \ \ldots \  \ldots \  \ldots \ \ldots \ \ldots \ \ldots \ \ldots \ \ldots \ \ldots \ \nonumber \\
    &g^{e}:& \sum_{j=0}^e \delta_j S^{(m)_k}_{e-j|(s)_k}[(\chi_c)_k] =0 . \label{ge}
 \end{eqnarray}
Second, we should take into account for influence of traceless constraints $L^{(i)}_{11}$ on the structure of vertices $\big|  V^{(p)}_c\rangle^{(m)_{(i)_p}}_{(s)_{(i)_p}}$, $\big|  \widetilde{V}{}^{(p)}_c\rangle^{(m)_{(i)_p}}_{(s)_{(i)_p}}$ for $p=3,4,...,e$.

The resolution of   (\ref{g1})  for cubic vertices leads to the system
\begin{eqnarray}
&& \mathcal{Q}(V^{(3)}_{c|{(i)_3}},\widetilde{V}^{(3)}_{c|{(i)_3}}) = \sum_{n=1}^3
Q^{(i_n)} \big|  \widetilde{V}^{(3)}_c\rangle
   +  Q^{(i_j)}\Big( \big|  V^{(3)}_c\rangle - \big|  \widetilde{V}^{(3)}_c\rangle\Big)=0
\label{g1operV3}   
\end{eqnarray}
 (for $j=1,2,3$) which particular solution for coinciding $V^{(3)}_{c|{(i)_3}} = \widetilde{V}^{(3)}_{c|{(i)_3}}$ has the usual form
but augmented  by the validity of spin and traceless conditions
\begin{eqnarray}\label{gencubBRSTc}
 &&  Q_c^{tot}
\big|  V^{(3)}_c\rangle^{(m)_{(i)_3}}_{(s)_{(i)_3}} =0,   \qquad     L^{(i_j)}_{11} \big|  V^{(3)}_c\rangle^{(m)_{(i)_3}}_{(s)_{(i)_3}} \ =\ 0, \\
 && \sigma^{(i)}_c\big|  V^{(3)}_c\rangle^{(m)_{(i)_3}}_{(s)_{(i)_3}}\ =\  \Big(s_i+\frac{d-2+\theta_{m_i,0}}{2}\Big)\big|  V^{(3)}_c\rangle^{(m)_{(i)_3}}_{(s)_{(i)_3}} .
\label{gencubBRSTc1}
\end{eqnarray} where $ Q_c^{tot}=\sum_{j=1}^k Q_c^{(j)}$. The vertex $ \big| V^{(3)}_c\rangle^{(m)_{(i)_3}}_{(s)_{(i)_3}} $ has a local representation:
\begin{equation}\label{xdep}
  \big |V^{(3)}_c\rangle^{(m)_{(i)_3}}_{(s)_{(i)_3}} = \prod_{i=1}^3 \delta^{(d)}\big(x -  x_{i}\big) V^{(3)}(x)
  \prod_{j=1}^3 \eta^{(j)}_0 |0\rangle , \ \  |0\rangle\equiv \otimes_{i=1}^e |0\rangle^{i}.
\end{equation}.

Let us verify  that the deformed  equations of motion   and  any representative $\big|\widetilde{\chi}^{(i)}_c\rangle_{s_i}$ from arbitrary  gauge orbit  $\mathcal{O}_{[1]|\chi^{(i)}_c}$,
\begin{equation}\label{goc1}
\mathcal{O}_{[1]|\chi^{(i)}_c} = \big\{\big|\widetilde{\chi}^{(i)}_c\rangle_{s_i} \big|\ \ \big|\widetilde{\chi}^{(i)}_c\rangle_{s_i} =\big|{\chi}^{(i)}_c\rangle_{s_i} + \delta_{[1]}|\chi^{(i)}_c\rangle_{s_i},  \, \forall \big|\Lambda^{(j)}_c\rangle_{s_j}\big\}
\end{equation} (for $ j=1,...,k $, with the properties $(\epsilon, gh)\big|{V}{}^{(3)}\rangle = (1,3)$)
  for interacting fields remains by traceless
after applying the deformed gauge transformations. Then, it is sufficient to find that
\begin{eqnarray}
&\hspace{-0.5em} &\hspace{-0.5em}   L^{(i)}_{11}\delta_{[1]}|\chi^{(i)}_c\rangle_{s_i}= L^{(i)}_{11}Q_c^{(i)}|\Lambda^{(i)}_c\rangle_{s_i} - g \int d\eta^{(i_1)}_0d\eta^{(i_2)}_0\Big( {}_{s_{i_1}}\langle
\Lambda^{(\{i_1)}_c\big|{}_{s_{i_2}}
   \langle \chi^{(i_2\})}_c\big|  \nonumber\\
   &\hspace{-0.5em} &\hspace{-0.5em}
\ \   \phantom{\delta_{[1]} \big| \chi^{(i)} \rangle_{s_i}}
\times L^{(i)}_{11}\big|{V}{}^{(3)}_c\rangle^{(m)_{i_2i}}_{(s)_{i_2i}} = 0, \label{cubgtrcc} \\
  &\hspace{-0.5em} &\hspace{-0.5em}  L^{(i)}_{11}\frac{\overrightarrow{\delta} S^{(m)_3}_{[1]C|(s)_3}}{\delta {}_{s_{i}}
   \langle \chi^{({i})}_c\big|}  = L^{(i)}_{11}Q_c^{(i)}|\chi^{(i)}_c\rangle_{s_i} + g
    \sum_{1\leq i_1<i_2\leq k}^{i_j\ne i} \hspace{-1.0em}
   \int  \prod_{j=1}^2 d\eta^{(i_j)}_0   {}_{s_{i_j}}\langle \chi^{(i_j)}_c \big|\nonumber\\
   &\hspace{-0.5em} &\hspace{-0.5em}
\ \   \phantom{\delta_{[1]} \big| \chi^{(i)} \rangle_{s_i}}\times
 L^{(i)}_{11} \big|  V^{(3)}_c\rangle^{(m)_{(i)_2i}}_{(s)_{i_2i}}= 0. \label{cubgtr1c}
\end{eqnarray}

We proved, that imposing of only traceless constraints on fields and gauge parameters (\ref{L11}) represents the necessary but not sufficient condition for the consistency
of deformed (on cubic level) Lagrangian dynamics.
Indeed, in this case the latter term in (\ref{cubgtrcc}),  (but not in (\ref{cubgtr1c})) does not vanish, and therefore the number of independent initial data (number of physical degrees of freedom) for deformed and undeformed cases are different. For the equations (\ref{cubgtr1c}) due to  completeness of the inner product in the total Hilbert space in question the solutions for constraints (\ref{L11}) generate  hermitian projectors $P^{(i_j)}_{m|11}$, $m=0,1$ on the subspaces of traceless field  and gauge vectors which extract from the vertex $\big|  V^{(3)}_c\rangle^{(m)_{(i)_3}}_{(s)_{(i)_3}}$ in (\ref{S[3]}) the traceless vertex $\big|  \overline{V}{}^{(3)}_c\rangle^{(m)_{(i)_3}}_{(s)_{(i)_3}} \equiv$ $\prod_{j=1}^3P^{(i_j)}_{0|11}\big|  V^{(3)}_c\rangle^{(m)_{(i)_3}}_{(s)_{(i)_3}}$. For the vertex incoming into deformed gauge transformations (\ref{gt1}) such traceless propety does not hold.

Obvious generalization of this requirement for the $p$-tic vertices  leads for $\big|  {V}^{(p)}_c\rangle^{(m)_{(i)_3}}_{(s)_{(i)_3}}$ in addition to the rest equations from (\ref{g2}), (\ref{ge}) to be by $L^{(i)}_{11}$-traceless solutions of the equations
\begin{equation}\label{ptrace}
   L^{(i_j)}_{11} \big|  V^{(p)}_c\rangle^{(m)_{(i)_p}}_{(s)_{(i)_p}} \ =\ 0,\ \ p=3,4,\ldots , e.
\end{equation}
  The $Q_c^{tot}$-closed traceless solutions for the  equations  (\ref{gencubBRSTc}), (\ref{gencubBRSTc1})
determines consistent cubic vertices for irreducible interacting  higher-spin fields with given masses and spins which Lagrangian formulation has  the same number of physical degrees of freedom as ones for the same free irreducible fields. This property is missed for the cubic vertices in \cite{BRST-BV3}.

\section{Relation to approach with complete BRST operator}
\label{sec:completeBRST}

A  cubic deformation for  $k$, $k\geq 3$ copies of Lagrangian formulations \cite{BRcub}, \cite{BRcubmass} with unconstrained  vectors
$|\chi^{(j)}\rangle_{s_j}$, zeroth and first level gauge parameters
$|\Lambda^{(j)l}\rangle_{s_j}$, $l=0,1$, is based on the  complete spin and  BRST operators which  include trace constraints, also  additional oscillators $b^{(j)}, b^{(j)+}$,
$\eta^{(j)+}_{11}$, $\eta^{(j)}_{11}$,  $\mathcal{P}^{(j)+}_{11}$, $\mathcal{P}^{(j)}_{11}$,
\begin{eqnarray}
\label{spin}
&\hspace{-0.5em}& \hspace{-0.5em} {\sigma}^{(j)}  =   \sigma^{(j)}_c + 2b^{(j)+}b^{(j)}+ h^{(j)}
 + 2(\eta^{(j)+}_{11}\mathcal{P}^{(j)}_{11} -\eta^{(j)}_{11}\mathcal{P}_{11}^{(j)+}),\\
 \label{Qtotsym}
&\hspace{-0.5em}& \hspace{-0.5em}  {Q}^{(j)} =Q^{(j)}_c
+
\eta^{(j)+}_{11}\widehat{L}{}^{(j)}_{11}+\widehat{L}{}^{(j)+}_{11}\eta^{(j)}_{11},  \ [b^{(i)}, b^{(j)+}]=\{\eta^{(i)+}_{11}, \mathcal{P}^{(j)}_{11}\} =\delta^{ij},    \\
&\hspace{-0.5em}& \hspace{-0.5em} \big(\widehat{L}{}^{(j)}_{11} ,\widehat{L}{}^{(j)+}_{11}\big) =  \big(
{L}^{(j)}_{11}+ (b^{(j)+}b^{(j)} +h^{(j)})b^{(j)}, \, {L}^{(j)+}_{11}+b^{(j)+} \big),
\label{Qtrace}
\end{eqnarray}
where $h = h(s)=-s - \frac{d-5-\theta_{m,0}}{2}$ and operator $K$ determining scalar product an hermitian conjugation: $AK=KA^+$, for $A\in\{\sigma, Q\}$.

The action with cubic vertex as a gauge theory of first-stage reducibility in a configuration space
$\mathcal{M}^{(s)_k}_{cl}$  parameterizing by basic  $\phi^{(i)}_{\mu(s)_i}$ and sets of auxiliary fields $\phi^{ (i)}_{1\mu({s_i}-1)},...$ of smaller rank, embedded in $|\chi^{(i)}\rangle_{s_i}$ of a Hilbert space, $\mathcal{H }^{(i)}_{tot}$ $=$ $\mathcal{H}^{(i)}\otimes \mathcal{H}^{(i)}{}'\otimes \mathcal{H}^{ (i)}_{gh}$, $i=1,...,k$
\begin{eqnarray}
\hspace{-1em}&& |\chi^{(i)}\rangle_{s_i}  =|\chi^{(i)}_c (b^+)\rangle_{s_i} + |\chi^{(i)}_{aux} \rangle_{s_i}, \ \ |\chi^{(i)}_{aux} \rangle\big|_{(\mathcal{P}_{11}^{+}= \eta_{11}^{+}=0)} \label{spinctotsym}
\end{eqnarray}
(with the similar structures for the gauge parameter $ \big| \Lambda^{(i)} \rangle_{s_i}$ and $ \big| \Lambda^{(i)1} \rangle_{s_i} =\mathcal{P}_{11}^{(i)+}\mathcal{P}_{1}^{(i)+}|\Xi^{(i)1}\rangle_{s_i-3}  $) takes the form
\begin{eqnarray}\label{S[n]}
   &\hspace{-0.5em} &\hspace{-0.5em} S_{[1]|(s)_k}[(\chi)_k]  =  \sum_{i=1}^{k} \mathcal{S}_{0|s_i}[\chi^{(i)}_{s_i}]   +
  g \hspace{-0.7em} \sum_{1\leq i_1<i_2<i_3\leq k} \hspace{-1.0em} \int \prod_{j=1}^3 d\eta^{(i_j)}_0  \Big( {}_{s_{i_j}}\langle \chi^{(i_j)}K^{(i_j)}\big|
  \\
  &\hspace{-0.5em} &\hspace{-0.5em}\phantom{S_{[1]|(s)_k}[(\chi)_k]  =} \times \big|  V^{(3)}\rangle^{(m)_{(i)_3}}_{(s)_{(i)_3}} +h.c. \Big)m\\
 &&  \hspace{-0.5em} \mathcal{S}_{0|s_i}[\chi^{(i)}_{s_i}] =  \mathcal{S}_{0|s_i}[\phi^{(i)},\phi^{(i)}_1,...]=
  \int d\eta^{(i)}_0 {}_{s_i}\langle\chi^{(i)}|
K^{(i)}Q^{(i)}|\chi^{(i)}\rangle_{s_i}.
\end{eqnarray}
It is invariant up to the first order in $g$ with respect to non-Abelian reducible gauge transformations with the same  vertices
\begin{eqnarray}
  && \delta_{[1]} \big| \chi^{(i)} \rangle_{s_i}  =  Q^{(i)} \big| \Lambda^{(i)} \rangle_{s_i} -
g \hspace{-1.0em}\sum_{1\leq i_1<i_2\leq k}\hspace{-0.5em} \int \hspace{-0.3em}\prod_{j=1}^2 d\eta^{(i_j)}_0  \Big[ {}_{s_{i_1}}\langle \chi^{(\{i_1)}
  K^{(i_1)}\big|
  \otimes   \label{cubgtr}\\
   &&
\ \   \phantom{\delta_{[1]} \big| \chi^{(i)} \rangle_{s_i}} \otimes {}_{s_{i_2}}\langle \Lambda^{(i_2\})}K^{(i_2\})}\big| {V}{}^{(3)}\rangle^{(m)_{(i)_2j}}_{(s)_{(i)_2j}}
, \nonumber\\
&& \delta_{[1]} \big| \Lambda^{(i)} \rangle_{s_i}  =  Q^{(i)} \big| \Lambda^{(i)1} \rangle_{s_i}
 -g \hspace{-1.0em}\sum_{1\leq i_1<i_2\leq k}\hspace{-0.5em} \int \hspace{-0.3em}\prod_{j=1}^2 d\eta^{(i_j)}_0  \Big[ {}_{s_{i_1}}\langle \chi^{(\{i_1)}
  K^{(i_1)}\big|
  \otimes    \label{cubggtr}\\
   &&
\ \   \phantom{\delta_{[1]} \big| \chi^{(i)} \rangle_{s_i}} \otimes {}_{s_{i_2}}\langle \Lambda^{(i_2\})}K^{(i_2\})}\big| {V}{}^{(3)}\rangle^{(m)_{(i)_2j}}_{(s)_{(i)_2j}}.
\nonumber
\end{eqnarray}
The validity of the operator equations at the  first order in $g$ for local vertex
\begin{equation}
\label{g1Lmod}
  Q^{tot}
\big|{V}{}^{(3)}\rangle^{(m)_{(i)_3}}_{(s)_{(i)_3}} = 0, \qquad \sigma^{(i)}\big|{V}{}^{(3)}\rangle^{(m)_{(i)_3}}_{(s)_{(i)_3}}\ =\ 0,
\end{equation}
guarantees (due to automatical inclusion of traceless conditions in $Q^{tot}$) that the deformed Lagrangian dynamics will determine the theory for irreducible interacting higher-spin fields with masses $(m)_{(i)_3}$ and spins $(s)_{(i)_3}$ with preservation the number of physical degrees of freedom.

The general solution of the equations (\ref{g1Lmod}) for a cubic vertex has been obtained in \cite{BRcub} for massless and \cite{BRcubmass} for massive irreducible fields as follows
\begin{eqnarray}\label{genvertex}
   \hspace{-1.0em} &\hspace{-1.0em}&\hspace{-1.0em}|{V}{}^{(3)}\rangle^{(m)_{(i)_3}}_{(s)_{(i)_3}} = |{V}{}^{M(3)}\rangle^{(m)_{(i)_3}}_{(s)_{(i)_3}}  +\hspace{-0.5em} \sum_{l=1;(j_{i_l}) >0}^{3;([s_{1_l}/2])}\hspace{-0.5em} U^{(s_{i_1})}_{j_{i_1}}U^{(s_{i_2})}_{j_{i_2}}U^{(s_{i_3})}_{j_{i_3}}|{V}{}^{M(3)}\rangle_{(s)_3-2(j)_{i_3}}\hspace{-0.1ex}.
   \end{eqnarray}
with new trace terms $U^{(s_{i_1})}$ (see e.g. \cite{Rcubmasless}) as compared to the constrained case. Note, for $k=3$ and vanishing "trace" oscillators $b^{(i)+}$, $ \mathcal{P}_{11}^{(i)+} $, $\eta_{11}^{(i)+}$
the vertex $|{V}{}^{M(3)}\rangle^{(m)_3}_{(s)_3}$ coincides with one $\big|  V^{(3)}_c\rangle^{(m)_3}_{(s)_3}$ (\ref{xdep}) obtained in \cite{BRST-BV3}.

\section{Conclusion}
\label{sec:conclusion}

In the present article, we have derived  the necessary and
sufficient conditions for cubic and $p$-tic ($p\geq 4$) vertices
obtained within approach with incomplete BRST operator to describe
consistent Lagrangian dynamics for irreducible interacting
totally-symmetric higher-spin fields in $d$-dimensional Minkowski
space subject to appropriate traceless constraints. It is shown that
imposing of only constraints above on field and gauge parameter
vectors (\ref{L11}) that form gauge-invariant content of  Lagrangian
formulation is insufficient to preserve the number of physical
degrees of freedom passing from free to interacting theory.
Additionally,  to above restrictions the set of incomplete BRST,
spin operators,  cubic vertices and traceless constraints must form
closed superalgebra (\ref{QsL11}), (\ref{gencubBRSTc}),
(\ref{gencubBRSTc1}). The constraints and cubic (also all $p$-tic)
vertices should supercommute. The solution for BRST-closed
traceless cubic vertices represents for constrained case  the
local threevector. At the same time  the application of the
approach with complete BRST operator (\ref{Qtotsym})  automatically
leads to local cubic vertices to be by BRST ($Q^{tot}$)-closed solution with given
spin  for (\ref{g1Lmod}) with the same interacting
higher spin fields, but depending  on  wider set of oscillators with
additional trace-inspired factors (\ref{genvertex}). The
correspondence among the cubic vertices in constrained and
unconstrained BRST approaches for the same irreducible  interacting
fields may be established after eliminating the auxiliary fields and
gauge parameters by partially fixing the gauge and using the
equations of motion, the vertices $|{V}{}^{(3)}\rangle$ will
transform to $|{\breve{V}}{}^{(3)}\rangle$ in a triplet formulation
of \cite{BRST-BV3} but to be by traceless, so that, up to
total derivatives, the vertices $\big|  \overline{V}{}^{(3)}_c\rangle$ and
$|{\breve{V}}{}^{(3)}\rangle$ (possibly being nonlocal) should
coincide.
\paragraph{Acknowledgements} The authors are grateful to organizers and participants of International
Workshop "Supersymmetries and Quantum Symmetries 2022" for
hospitality and useful discussions of the presented results. The
work was partially supported by the Ministry of Education of Russian
Federation, project No QZOY-2023-0003.
%


\end{document}